\documentclass[%
reprint,
superscriptaddress,
 amsmath,amssymb,aps,
 twocolumn,
 citeautoscrip]{revtex4-1}
\usepackage{graphicx}
\usepackage{dcolumn}
\usepackage{bm}
\usepackage{color,soul}
\usepackage[colorlinks=true, linkcolor=blue, citecolor=blue,urlcolor=blue,breaklinks=true]{hyperref}
\usepackage{soul}
\usepackage{comment}
\newcommand{\ket}[1]{\left|{#1}\right\rangle}
\newcommand{\bra}[1]{\left\langle{#1}\right|}

\begin{document}

\title{Heralded quantum steering over a high-loss channel}

\author{Morgan M. Weston}
\affiliation{Centre for Quantum Dynamics and Centre for Quantum Computation and Communication Technology, Griffith University, Brisbane, Queensland 4111, Australia}
\author{Sergei Slussarenko}
\affiliation{Centre for Quantum Dynamics and Centre for Quantum Computation and Communication Technology, Griffith University, Brisbane, Queensland 4111, Australia}
\author{Helen M. Chrzanowski}
\affiliation{Centre for Quantum Dynamics and Centre for Quantum Computation and Communication Technology, Griffith University, Brisbane, Queensland 4111, Australia}
\affiliation{Clarendon Laboratory, University of Oxford, Parks Road, Oxford OX1 3PU, UK}
\author{Sabine Wollmann}
\affiliation{Centre for Quantum Dynamics and Centre for Quantum Computation and Communication Technology, Griffith University, Brisbane, Queensland 4111, Australia}
\affiliation{Quantum Engineering Technology Labs, H. H. Wills Physics Laboratory and Department
	of Electrical and Electronic Engineering, University of Bristol, BS8 1FD, UK}
\author{Lynden K. Shalm}
\affiliation{National Institute of Standards and Technology, 325 Broadway, Boulder, Colorado 80305, USA}
\author{Varun B. Verma}
\affiliation{National Institute of Standards and Technology, 325 Broadway, Boulder, Colorado 80305, USA}
\author{Michael S. Allman}
\affiliation{National Institute of Standards and Technology, 325 Broadway, Boulder, Colorado 80305, USA}
\author{Sae Woo Nam}
\affiliation{National Institute of Standards and Technology, 325 Broadway, Boulder, Colorado 80305, USA}
\author{Geoff J. Pryde}
\affiliation{Centre for Quantum Dynamics and Centre for Quantum Computation and Communication Technology, Griffith University, Brisbane, Queensland 4111, Australia}

\date{\today}
\begin{abstract}
	Entanglement is the key resource for many long-range quantum information tasks, including secure communication and fundamental tests of quantum physics. These tasks require robust verification of shared entanglement, but performing it over long distances is presently technologically intractable because the loss through an optical fiber or free-space channel opens up a detection loophole. We design and experimentally demonstrate a scheme that verifies entanglement in the presence of at least $14.8\pm0.1$ dB of added loss, equivalent to approximately $80$ km of telecommunication fiber. Our protocol relies on entanglement swapping to herald the presence of a photon after the lossy channel, enabling event-ready implementation of quantum steering. This result overcomes the key barrier in device-independent communication under realistic high-loss scenarios and in the realization of a quantum repeater.
\end{abstract}
\maketitle
A reliable method to send quantum information---from Alice to Bob, say---over long distances is to teleport it, using entanglement shared by the two remote parties~\cite{book_nielsen00,horodecki09}. This entanglement resource could alternatively be used for generating secure correlated randomness between Alice and Bob, or efficiently completing shared computational tasks~\cite{rev_ralph09}, or testing nonlocality and quantum mechanics in new regimes such as when the parties are in different relativistic reference frames~\cite{rideout2012,alsing2012}. All of these applications reach their potential only when entanglement is distributed over a long distance. Photons are excellent carriers of the quantum information, being a quantum version of the optical encodings used in existing long-distance classical telephony and data networking. However, in the quantum regime, attenuation (photon loss) is very destructive, because the noise added by this process corrupts the entanglement. Thus, the maximum length of a quantum communication channel is restricted by propagation loss and environmental contamination. 

Importantly, the most secure quantum communication approaches---device-independent protocols~\cite{acin06,branciard12,comandar16}---and the most robust tests of nonlocality and quantum mechanics require verification of entanglement with stringent conditions on the amount of tolerable loss. For example, complete entanglement verification, through a violation of a Bell inequality, has recently been  performed with the three main loopholes~\cite{larsson14} closed simultaneously~\cite{shalm15, giustina15,hensen15}. Though these results  represent a  significant advance, practical limitations  remain  in exploiting these tests in a realistic long-distance scenario.  Inevitable losses through  any fibre or free-space  channel open the detection loophole~\cite{pearle70} for standard  photonic implementations, forbidding  a robust test even when the postselected measurement correlations  are strong enough to violate a Bell inequality.

Here, to overcome the effect of loss in quantum channel, we adopt an \textit{event-ready} approach~\cite{zukowski93}. The key idea is  to record an additional  heralding  signal that indicates whether the quantum state under investigation was successfully shared between Alice and  Bob---that is,  whether the particles are ready to be used in the verification protocol. By conditioning the validity of the protocol trial on this  heralding  signal, failed distribution events are excluded  beforehand  from being used in the test. We use entanglement swapping~\cite{pan98,kaltenbaek09,jin15} to realize an event-ready scheme, allowing us to perform entanglement verification, detection loophole-free, over lossy quantum communication channels.  Our approach also represents a central element of a more complex quantum repeater architecture, which may be used to overcome loss in very large networks. 

We use an alternative test of nonlocality, quantum steering (also called Einstein-Podolsky-Rosen steering). Steering is an asymmetric protocol where one party, Bob, trusts quantum mechanics to describe his own measurements, while no assumptions are made about Alice, the other, untrusted, party~\cite{wiseman07,saunders10}. The test may be satisfied by using entanglement to steer the state of a distant quantum system by local measurements on its counterpart. The nonlocal correlations verified by steering both guarantees shared entanglement~\cite{saunders10} and may be configured, with certain conditions, to implement one-sided device-independent QKD~\cite{branciard12}.  

\begin{figure}
	\includegraphics{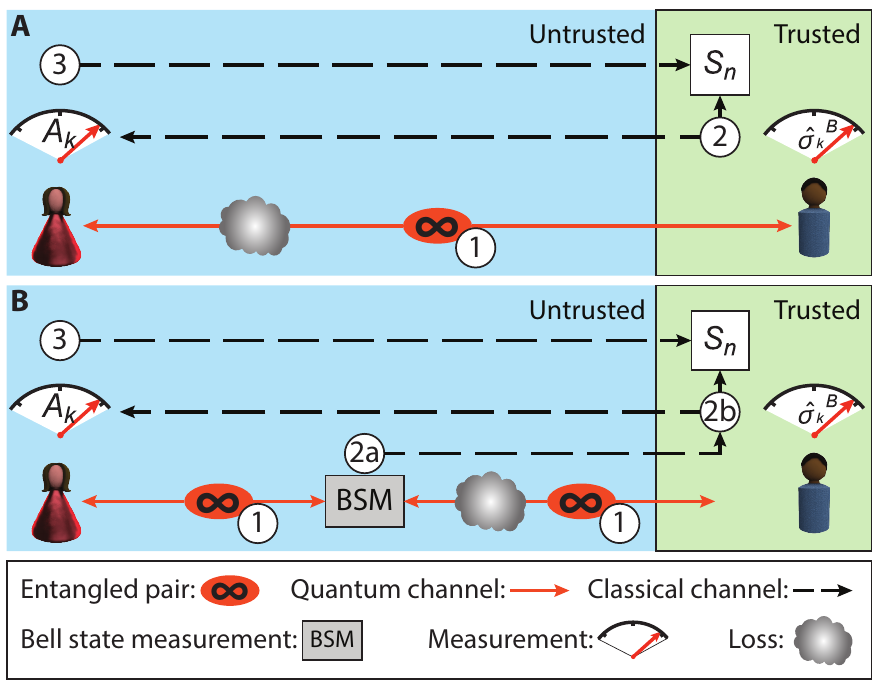}
	\caption{Conceptual representation of the  quantum steering protocols. The blue background denotes untrusted channel components that belong to Alice and the green background denotes the trusted side, Bob. ({\bf A}) Conventional steering: (1) Alice prepares a pair of photons and sends one of them to Bob; (2) Bob announces his measurement setting, $k$, from a predetermined set of $n$ observables; (3) Bob records his measurement outcome $\hat{\sigma}^B_k$  and Alice declares her result $A_k$; steps (1) to (3) are iterated to obtain the steering parameter $S_n$. ({\bf B}) Heralded quantum steering protocol. Bob uses a classical signal from a successful Bell state (BSM) measurement (2a) to herald the presence of Alice's photon after the lossy channel, ignoring all the trials when the BSM was not successful.}
	\label{concept}
\end{figure}

In quantum steering,  Alice's task is to convince Bob that she can influence his quantum measurement outcome for any choice of measurement setting that Bob provides to her. 
The formal steering protocol is shown in Fig.~\ref{concept}A. Bob’s choice of measurement setting, labeled $k$, is chosen from a predetermined set of $n$ observables. Bob’s $k$-th measurement setting corresponds to the Pauli observable $\hat{\sigma}^B_k $ for $ k \in \{1,..,n\} $. We make no assumptions about what Alice is doing and thus represent her results as $A_k  \in \{-1,1 \}$ Steps 1 to 3, from Fig.~\ref{concept}A, are iterated to obtain the average correlations between Alice’s and Bob’s
results, known as the steering parameter~\cite{saunders10} 
\begin{equation}\label{S_n}
S_n \equiv \dfrac{1}{n} \sum^n_{k=1} \langle A_k \hat{\sigma}^B_k \rangle.
\end{equation}
If $S_n$ is larger than a certain bound $C_n$~\cite{saunders10}, then Alice has successfully demonstrated quantum steering. 
Correct timing of these events is necessary to close  the locality loophole~\cite{wittmann12} and Bob must have truly random measurement  choices in order to close the freedom of choice loophole. However, the focus of this work is on transmission loss.

\begin{figure*}
\includegraphics{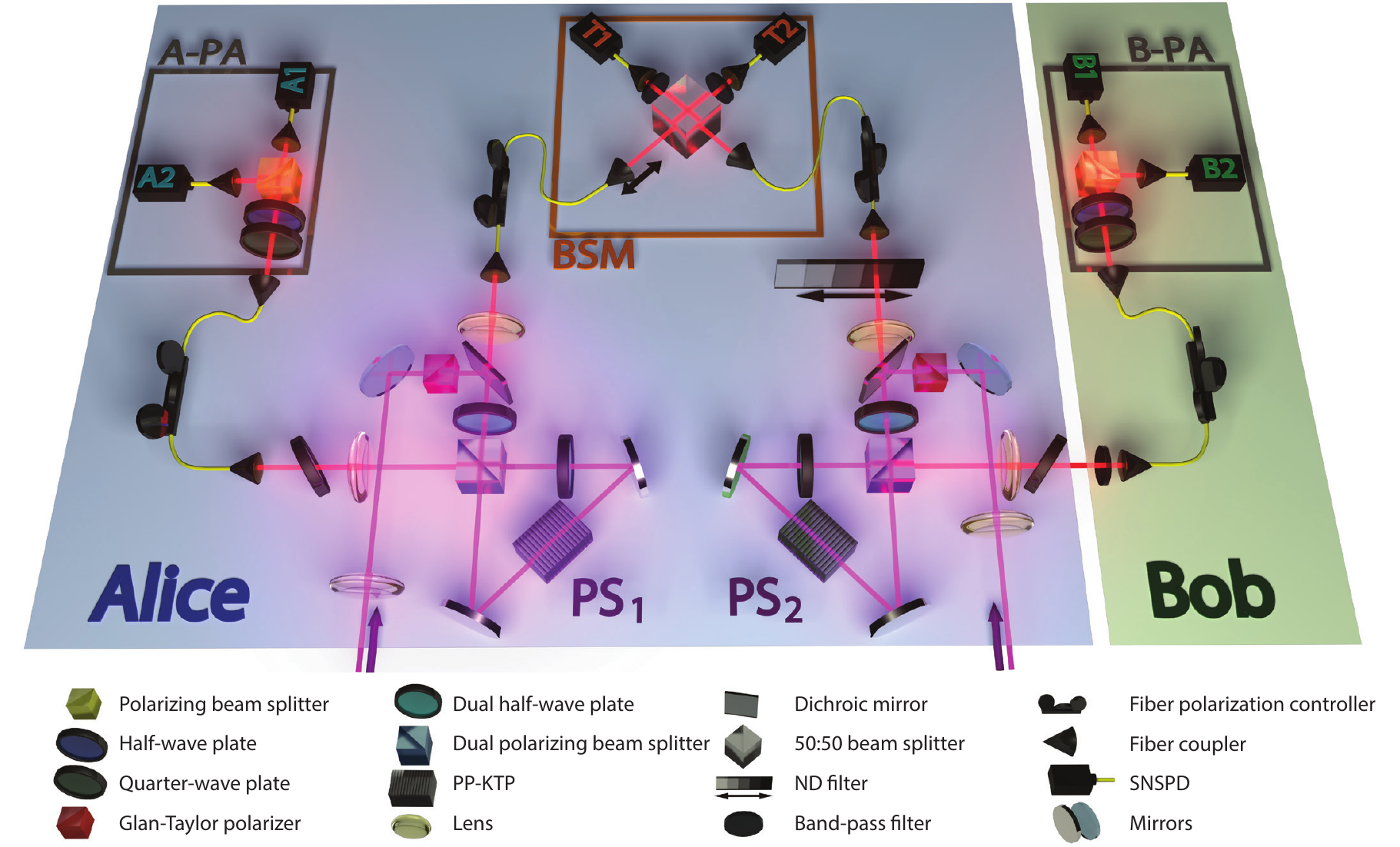}
\caption{ Experimental setup. Two group-velocity-matched sources, $\mathcal{PS}_1$ and $\mathcal{PS}_2$, are pumped by a  mode-locked femtosecond Ti:sapphire laser to generate two polarization-entangled photon pairs at $1570~\mathrm{nm}$, in the $\ket{\Psi^-}$ state. Blue and green backgrounds outline the untrusted and trusted sides respectively. (All untrusted elements are grouped with Alice, even if they are not in her ``lab'' in practice.) A-PA and B-PA are the polarization analyzer (tomography) stages of Alice and Bob, and BSM is the Bell state measurement gate, composed of a nonpolarizing 50:50 beam splitter.  A variable neutral density (ND) filter is used in the Alice's arm of $\mathcal{S}_2$ to introduce the channel loss, $\mathcal{L}$.  Eight-nm band-pass filters (BP) were placed in the path of the photon going to B-PA and after beam splitter (BS) increasing the singlet state fidelity, while maintaining Alice's high heralding efficiency. For the conventional steering measurement, Alice's arm of $\mathcal{PS}_2$ was directly connected to the A-PA stage through the fiber, bypassing the BSM gate and $\mathcal{PS}_1$. SNSPD, superconducting nanowire single-photon detector; PP-KTP, periodically poled potassium titanyl phosphate.}
\label{EXPsetup}
\end{figure*}
\renewcommand{\thefigure}{\arabic{figure}}

A dishonest Alice---or an eavesdropper controlling Alice's apparatus---may attempt to use the fair sampling assumption (i.e.\ an open detection loophole) to cheat, by hiding incompatible measurement results~\cite{pearle70}.
She may mimic perfect correlations of a maximally entangled state, and Bob has no way to determine whether a lack of measurement outcome announcement by Alice is due to genuine qubit loss or cheating. To prevent cheating, Bob requires Alice to announce her measurement result at least a certain fraction $\varepsilon$ of trials, which we call Alice's heralding efficiency. When the entanglement verification is performed over long distances, the additional loss in Alice's channel will inevitably reduce the heralding efficiency below an acceptable value required for loophole-free entanglement verification.
 
The generalised steering bounds~\cite{bennet12}, which take into account Alice's heralding efficiency, allow detection-loophole-free quantum steering in presence of arbitrarily high loss in the untrusted quantum channel. However, guaranteed success for very high channel loss relies on the use of perfect pure entangled states and an infinite number of measurement settings, which is unrealistic in real-world scenarios.

Implementing an event-ready entanglement verification scheme allows us to herald the presence of the qubit in Alice's arm, increasing her effective heralding efficiency. In principle, this improved heralding could be realised in one of several ways, broadly including quantum-non-demolition-style measurements such as entanglement swapping \cite{pan98,kaltenbaek09,jin15}, distillation techniques such as noiseless linear amplification~\cite{kocsis13,ulanov15}, and photonic qubit precertification~\cite{meyerscott16}. For the levels of loss considered here, we favour entanglement swapping, owing to its comparatively low resource overhead and high success rates. If entanglement swapping is performed with spontaneous parametric donwconversion (SPDC) sources, as here, the squeezing parameter of those sources must be chosen carefully to control the effect of multiphoton events (see Table~\ref{table_Powers}, Methods, and fig.~\ref{highorders}). 

\begin{table}
	\centering
	\caption{Experimental parameters and rates. The pump power $P$, approximate counting time and total number of fourfold coincidence counts measured for different amount of loss, $\mathcal{L}$.}\label{table_Powers}
	\begin{tabular}{|c|c|c|c|c|}
		\hline 
		$\mathcal{L}$ & $P(\mathcal{PS}_1)$& $P(\mathcal{PS}_2)$& Count time& Fourfold \\ 
		(dB)&(mW) &(mW)&(hours)&(counts)\\
		\hline 
		$7.7$ & $100$ & $50$ & $8.3$ & 730\\ 
		\hline 
		$11.3$ & $90$ & $40$ & $21.6$ & 549 \\ 
	\hline 
	$14.8$ & $75$ & $40$ & $98.5$ & 594 \\ 
	\hline 
\end{tabular} 
\end{table}
The effective increase in heralding efficiency obtained with the entanglement swapping step allows us to avoid the detection loophole without assuming the honesty of Alice. Crucially, if the protocol is run in a fully time-ordered mode, as illustrated in fig.~\ref{time}, Alice is forced to announce over a classical channel when the teleportation has been successful, and thus declare which subset of measurement runs should be used to verify the entanglement, before Bob announces his measurement settings. This prevents her exploiting the non-determinism of the swapping operation to introduce a loophole---announcing a false outcome of the entanglement swapping measurement gains her no advantage. Bob can then proceed with the steering verification protocol. 

We performed our experiment using two polarization-entangled photon pairs generated by separate high-heralding-efficiency sources~\cite{weston16}, $\mathcal{PS}_1$ and $\mathcal{PS}_2$ 
(see Fig.~\ref{EXPsetup} and Methods). The photon pairs were prepared in $\ket{\Psi^-}=(\ket{HV}-\ket{VH})/\sqrt{2}$, where $\ket{H}$ and $\ket{V}$ denote horizontal and vertical polarizations, respectively. The signal photon from source $\mathcal{PS}_2$ was sent to Bob for polarization analysis (B-PA) measurement, whereas the remaining photon from the same pair was sent through a variable-loss channel towards Alice. 

\begin{figure*}
\includegraphics{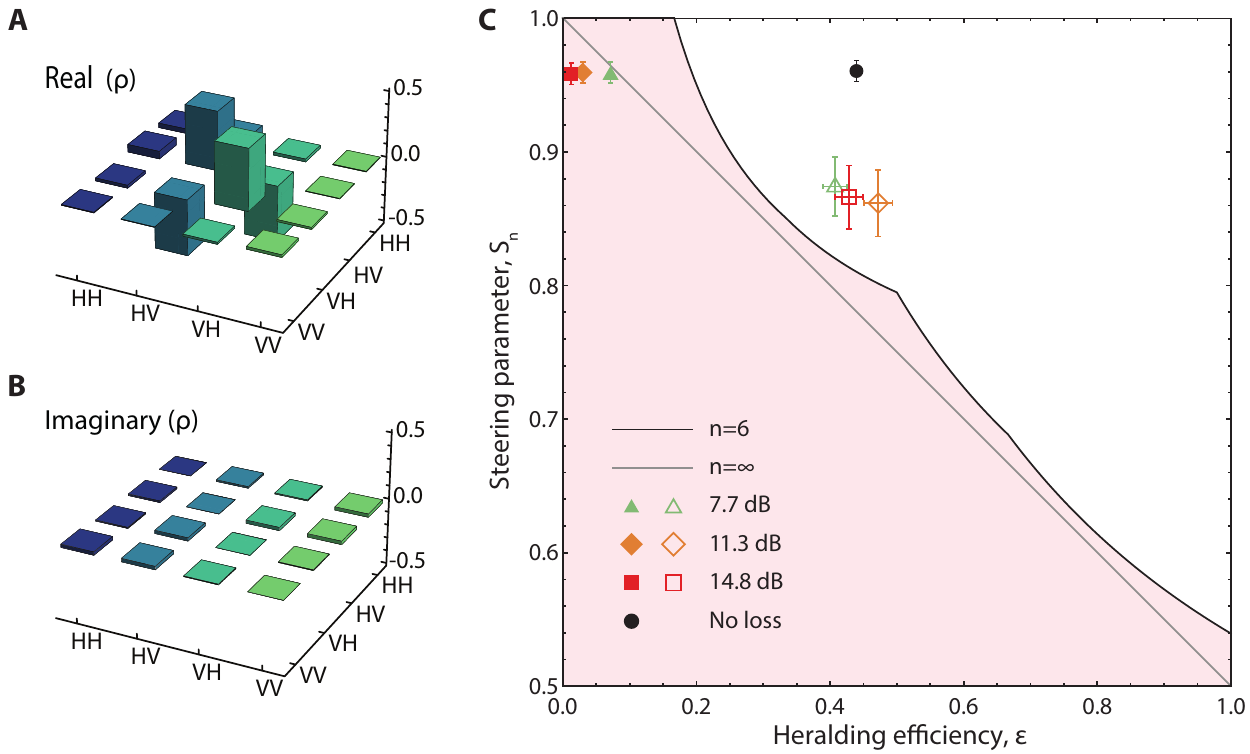}
\caption{Experimental results. ({\bf A}) Real  and ({\bf B}) imaginary parts of the 
reconstructed density matrix $\rho$ of the entanglement-swapped two-photon state with no additional loss applied to the quantum channel. ({\bf C}) Quantum steering measurement results for different amount of channel loss. Black and gray lines are, respectively, the $C_6(\varepsilon)$ and $C_\infty(\varepsilon)$ steering bounds from the study of Bennet \textit{et al.}~\cite{bennet12}, and red background highlights the  region where detection loophole-free steering with $n=6$ measurements fails. The black circle, green triangles, yellow diamonds, and red squares mark the steering results achieved in presence of $0$, $7.7\pm0.1$, $11.3\pm0.1$ and $14.8\pm 0.1~\mathrm{dB}$ of added channel loss, respectively. Filled markers correspond to steering parameters measured with the conventional steering protocol. Empty markers correspond to the heralded quantum steering results, each calculated from at least $500$ fourfold coincidence counts (Table~\ref{table_Powers}).}
\label{results}
\end{figure*}

In the first experiment, only source $\mathcal{PS}_2$ was used, and Alice's channel contained no added loss. Alice directly received her photon from $\mathcal{PS}_2$ into her polarisation analyzer (A-PA)---no entanglement swapping was used. We performed the steering protocol with $n=6$ measurement settings.
Note that as the number of measurement settings is increased in steering protocols, the observation of steering is slightly more tolerant to loss~\cite{bennet12}, but with diminishing returns. At the same time, the deleterious effect of measurement-setting errors grows with increasing $n$. The choice of $n=6$ provides a suitable balance between these effects. We observed detection loophole–free quantum steering with a steering parameter of of $S_6=0.960\pm0.008$ and Alice's heralding efficiency of $\varepsilon=0.4395\pm0.0003$,  violating the $C_n$ bound by $18$ standard deviations (SDs).

In the second experiment, a variable channel loss $\mathcal{L}$ was added between the two parties by using gradient ND filter (see Fig.~\ref{EXPsetup}). Using only source $\mathcal{PS}_2$, we measured the steering parameter $S_6$ and Alice's effective heralding efficiency for various levels of added channel loss. 
The results are shown in Fig.~\ref{results}C.  
With the addition of even $7.7\pm0.1~\mathrm{dB}$ of channel loss, the heralding efficiency dropped below the $C_\infty(\varepsilon)$ bound, forbidding secure  quantum steering even in the limiting theoretical case of infinite measurement settings. 

In the third experiment, we added source $\mathcal{PS}_1$ and the entanglement swapping step, realized through a partial BSM, to herald Alice's photon (see Methods for details). An additional coincidence detection signal from the two BSM detectors herald a successful swapping operation.  Alice's effective heralding efficiency at A-PA is now defined as the probability of detecting a four-photon coincidence from A-PA, B-PA and the triggers at the BSM, given that a three-photon coincidence was detected from B-PA and triggers at the BSM. The final shared state $\rho$, with no added loss, was determined using quantum state tomography~\cite{white07} and is shown in Fig.~\ref{results}(A and B). It had a singlet Bell state fidelity $\mathcal{F} = \bra{\Psi^{-}} \rho \ket{\Psi^{-}} = \left( 91\pm3\right) \%$ which is comparable with the best value previously reported in entanglement swapping \cite{wu13}. Because $\mathcal{PS}_1$ does not have unit heralding efficiency, the entanglement swapping does not produce deterministic arrival of a photon at Alice's polarization measurement. However, it increases this probability to a level compatible with demonstrating detection loophole–free steering: Her conditional heralding efficiency was recovered to $\varepsilon\sim0.45$ (Fig.~\ref{results}C).

Measured steering parameters of $0.874\pm0.022$, $0.862\pm0.022$, and
$0.866\pm0.024$ and heralding efficiencies of $0.41\pm0.02$, $0.47\pm0.02$,
and $0.43\pm0.02$, achieved in the presence of $7.7\pm0.1$, $11.3\pm0.1$, and
$14.8\pm0.1$ dB of added loss, respectively, correspond to a successful
violation of the $C_6(\varepsilon)$ bound by at least $2.2$ SDs. A channel-added loss of $14.8\pm 0.1~\mathrm{dB}$ is equivalent to $74-82 ~\mathrm{km}$ of telecom optical fiber, assuming fiber loss of $0.18$ to $0.2~\text{dB/km}$. It is worth noting that  Alice's total channel loss, including the loss due to optical components in the BSM gate but excluding the detector efficiency, amounts to $20.0\pm0.1~\mathrm{dB}$, which is equivalent to at least $100~\mathrm{km}$ of telecom fibre. 

As seen from Fig.~\ref{results}C, we have not observed any degradation in the measured steering parameter or heralding efficiency while increasing the amount of channel loss. This result suggests that the protocol is not limited to the demonstrated $14.8\pm 0.1~\mathrm{dB}$ of added ($20.0\pm0.1~\mathrm{dB}$ in total) loss,  and that achieving heralded quantum steering with higher values of loss is possible. Although the protocol can theoretically hold for arbitrarily high loss, increasing the loss significantly reduces the count rates (Table~\ref{table_Powers}). Increasing loss will eventually result in unrealistic count times and in spurious coincidence rates caused by dark and background counts becoming comparable to real coincident photon detections. The point (that is, degree of loss) at which this happens is highly dependent on details of detector performance; because this is an area of rapid development in the community, we think that it is inadvisable to provide a specific numerical estimate at this time. Because the main breakthrough of our work is closing the detection loophole over a high-loss channel, we did not implement randomized choice of measurement settings and time order of detection events in any of the steering protocol experiments.  

Our heralded quantum steering protocol is the first demonstration of detection-loophole-free entanglement verification over a high loss channel. The ability to keep the quantum steering detection loophole closed with total losses of at least $20\pm 0.1~\mathrm{dB}$, and potentially higher, opens many new possibilities for security in long range transmission through optical fibre, free space, or between earth and satellite.
With additional assumptions, it has been previously been shown how to make a measurement-device-independent version of the steering protocol~\cite{kocsis15} and how to turn quantum steering into a one-sided device-independent quantum key distribution scheme~\cite{branciard12}. The result we achieved is a considerable step towards the implementation of secure quantum communication, and represents a single step quantum relay, a crucial component for future quantum repeaters. 

\section*{Methods}
\subsection{Photon sources and characterization}
The heralded quantum steering protocol relies on pure and indistinguishable entangled states. Conventional spontaneous parametric downconversion (SPDC) sources require narrowband ($\text{FWHM}<2~\text{nm}$) filters to erase the spectral distinguishability. These filters significantly decrease the heralding efficiency, making a detection-loophole-free steering inequality violation impossible, unless other sources of noise and imperfection in the state are quite small, which is impractical for future\ practical application. To circumvent this obstacle, we developed a new type of SPDC photon source~\cite{weston16}. Our sources work at the  group  velocity matching (GVM) condition to generate frequency uncorrelated photon pairs, removing the necessity of harsh spectral filtering. 

We used two  such  polarization-entangled photon pair sources in the Sagnac interferometer configuration ($\mathcal{PS}_1$ and $\mathcal{PS}_2$ in Fig.~\ref{EXPsetup}), pumped by a mode-locked Ti:sapphire laser with $81~\text{MHz}$ repetition rate, $785~\text{nm}$ wavelength and $5.35~\text{nm}$ FWHM bandwidth. In each source, SPDC from a nonlinear periodically poled KTP crystal produced photon pairs with $1570~\text{nm}$ wavelength and $\approx15~\text{nm}$ FWHM bandwidth.  
The Sagnac configuration provided polarization entanglement, while the pump and crystal parameters were selected to remove frequency correlations. The focusing and collection Gaussian beam modes were
optimized to increase the heralding efficiency of the source, according to Weston \textit{et al.\ }\cite{weston16}. 
Together with high efficiency  superconducting  nanowire single photon detectors (SNSPD)~\cite{marsili13} we achieved heralding efficiencies of  $0.47 \pm 0.02$  for each of our  sources,  while maintaining the necessary high fidelity of entanglement swapping.

The entangled states produced by $\mathcal{PS}_1$ and $\mathcal{PS}_2$ were individually characterized via polarization state tomography. For each state we used either A-PA or B-PA for one of the photon measurements. To measure the other photon in a pair, an additional polarization measurement stage inserted in the BSM gate part of the setup (not shown in Fig.~\ref{EXPsetup}). The fidelity with the maximally-entangled singlet Bell state, measured in this way, was $\mathcal{F}=(97.2\pm0.3)\%$ for $\mathcal{PS}_1$, where both photons were filtered with $8~\mathrm{nm}$ band-pass filters, and $\mathcal{F}=(98.2\pm0.3)\%$ for $\mathcal{PS}_2$, where the spectral filtering was applied only to the photon passing through the BSM part of the apparatus. Although the source needs very little spectral filtering in principle, this moderate filtering was employed to overcome reductions in the fidelity caused by wavelength-dependent effects introduced by some optical components.  

\subsection{Entanglement swapping}
The entanglement swapping works as follows. The photon Alice receives from $\mathcal{PS}_2$ and one of the photons from $\mathcal{PS}_1$ are input to a BSM gate, where they interfere nonclassically on a 50:50 nonpolarizing beam splitter (the remaining photon from $\mathcal{PS}_1$ was sent to Alice’s polarization
analyzer). A coincidence detection signal from the two BSM detectors labels a successful projection onto the $\ket{\Psi^-}$ state, heralding a successful swapping operation. High-visibility Hong-Ou-Mandel (HOM) interference is required to perform the swapping operation with high fidelity.
This was achieved while maintaining high entangled-state fidelity and high heralding efficiency, thanks to our high-performance sources~\cite{weston16} and high-efficiency, low-noise SNSPDs~\cite{marsili13}.

We measured HOM interference visibility in the BSM gate in order to characterize the indistinguishability of the photons from $\mathcal{PS}_1$ and $\mathcal{PS}_2$. One photon from each maximally entangled pair was sent into BSM, while the remaining photons of each pair were sent into A-PA and B-PA gates and projected into the $\{\ket{H},\ket{V}\}$ basis. We used $8~\mathrm{nm}$ band-pass filters at the output of the BSM stage and on the heralding photon of the $\mathcal{PS}_2$ photon pair. No polarization optics were used inside the BSM gate. We observed visibilities of  $(90\pm 3)\%$ for $\ket{V}$ polarized photons and $(99\pm4)\%$ for $\ket{H}$ polarized photons. We attribute the lower visibility value to the residual polarization distinguishability of interfering photons, arising from the non-symmetric ($47:53$) splitting ratio of our BS for vertical polarization and, more significantly, from the lack of spectral flatness of optical coatings of optical components over the wide frequency band of our photons.

\subsection{Channel loss}
Alice's added channel loss, $\mathcal{L}$, was implemented  by using two variable neutral density filters, whose transmission was characterized separately using a $1570~\mathrm{nm}$ diode laser. $8~\mathrm{nm}$ BP filters introduced an additional $3.5\pm0.1~\mathrm{dB}$ of loss, and the loss due to optical  components and fiber coupling of the BSM gate was measured to be $1.7\pm0.1~\mathrm{dB}$. Together with $\mathcal{L}=14.8\pm 0.1~\mathrm{dB}$ of maximum added loss, the highest total loss applied to the channel was $20\pm 0.1~\mathrm{dB}$, excluding the nonunit quantum efficiency of the SNSPDs.

\subsection{High order SPDC pair generation}
The pumping powers $P(\mathcal{PS}_1)$ and $P(\mathcal{PS}_2)$ of our sources were kept below $100~\mathrm{mW}$ in order to keep negligible the impact of high-order pair production on the independent HOM interference and state quality~\cite{weston16,fulconis07}. In the presence of added loss, the fractional contribution of high order terms from $\mathcal{PS}_1$ increases. However, the probability of generating a photon pair from each source is comparable with the probability of generating two photon pairs from $\mathcal{PS}_2$. The latter events produce false heralding coincidences, significantly decreasing Alice's heralding efficiency (see Fig.~S2). Efficient heralding of entanglement swapping requires that the number of photons from one side (e.g.\ source $\mathcal{PS}_2$) is significantly lower than the number of photons from the other side (source $\mathcal{PS}_1$). With the increased channel loss, this condition is satisfied automatically even for equal raw  brightness of $\mathcal{PS}_1$ and $\mathcal{PS}_2$. Nevertheless, careful selection of appropriate pump powers for these sources is still required.  We found that as the loss was increased, $P(\mathcal{PS}_1)$ had to be decreased in order to keep the swapped state quality high, and $P(\mathcal{PS}_2)$ had to be matched accordingly to maintain the heralding efficiency. The pump powers chosen for different levels of added loss are shown in Table~\ref{table_Powers}.

\subsection{Experimental uncertainties}

The measurement uncertainties for the quantum steering parameters comprise heralding efficiency uncertainty and steering paramenter uncertainty, denoted by horizontal and vertical error bars, respectively, in Fig.~\ref{results}C. The latter takes into account both systematic measurement error and Poissonian photon counting noise. The systematic error contribution occurs due to the imperfections in optical components of Bob's measurement apparatus, which could result in an overestimate of the steering parameter. We used the systematic error estimation procedure developed by Bennet \textit{et al.}~\cite{bennet12}. This procedure assumes a perfect entangled state, and attributes any deviation from $S_n=1$ to the imperfection of  Bob's measurement apparatus.  Such an approach overestimates the steering parameter systematic uncertainty, when applied to our experimental data, where $S_n<1$ is known to arise predominantly from imperfect entangled states and imperfect entanglement swapping. The uncertainties in state parameters derived from quantum state tomography were calculated through standard error propagation techniques applied to a distribution of reconstructed density matrices arising from a Monte Carlo calculation which samples from Poissonian distributions of photon counts.

%

\vspace{1 EM}
\noindent\textbf{Acknowledgements}
 We thank J. Ho for help with SNSPDs and R. B. Patel for assistance with  data acquisition code.
Part of this work was supported by Australian Research Council grant DP140100648 and part of this work was supported by Australian Research Council grant CE110001027. M.M.W. and S.W. acknowledge support by the Australian Government Research Training Program scholarship.
\vspace{1 EM}

\noindent\textbf{Author contributions}

\noindent G.J.P. conceived
the idea. H.M.C. and G.J.P. designed the experiment. M.M.W., S.S., and H.M.C. constructed
and carried out the experiment with help and supervision from G.J.P. M.M.W. and S.S.
analyzed the data. S.W. assisted in early stages of development. L.K.S. assisted with source
development. V.B.V., M.S.A., and S.W.N. developed high-efficiency SNSPDs. All authors
discussed the results and contributed to the manuscript

\vspace{1 EM}

\noindent\textbf{Competing interests}

\noindent S.W.N. and
V.B.V. are inventors on a patent related to this work filed by the National Institute of
Standards and Technology (publication no. US9240539 B2, filed 24 April 2013). All other
authors declare that they have no competing interests.
\clearpage
\renewcommand{\thefigure}{S\arabic{figure}}
\setcounter{figure}{0}  
\begin{figure*}[t!]
	\includegraphics{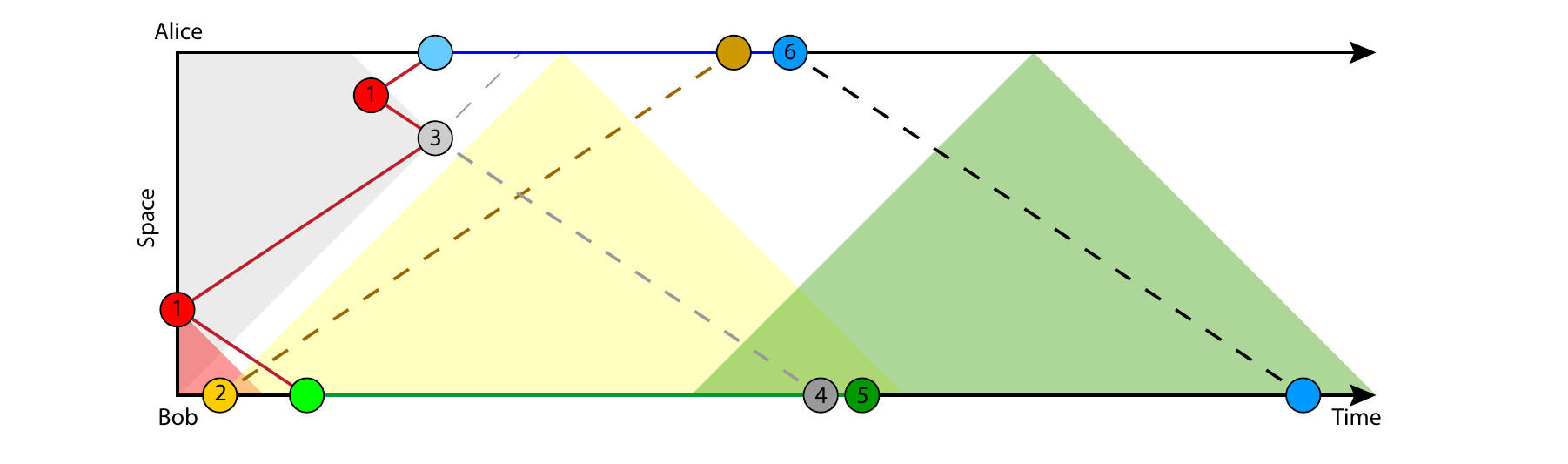}
	\caption{Proposed space-time diagram of the heralded steering protocol, illustrating the conditions to close the locality and freedom-of-choice loopholes. Red circles (1) illustrate the generation of photon pairs which are sent to Alice, Bob and the BSM gate. In order to close the freedom-of-choice loophole, the quantum random number generation of measurement settings, yellow circle (2), should be space-time separated from the photon generation, i.e.\ located within the red area on the diagram. The BSM, represented by gray circle (3), should occur before Alice, or the BSM gate may, in principle, gain access to measurement settings (yellow area). Following B. Wittmann \textit{et al.}~\cite{wittmann12}, the setting information independence is then enforced by Bob's BSM detection (4) and measurement (5) happening within the yellow shaded area, which denotes space-time separation from  the event marking the time at which Alice could  in  principle know measurement
		settings. The outcome-independence loophole is then closed by enforcing space-time separation (green area) between the  event of  Alice's  outcome  report, blue  dot (6), and Bob's measurement event (5). Both quantum (red lines) and classical (gray and black dashed lines) communication signals  are assumed to travel at $2/3$ of speed of light. Note that time ordering and measurement choice randomization was not implemented in this work.}
	\label{time}
\end{figure*}
\begin{figure*}
	\includegraphics{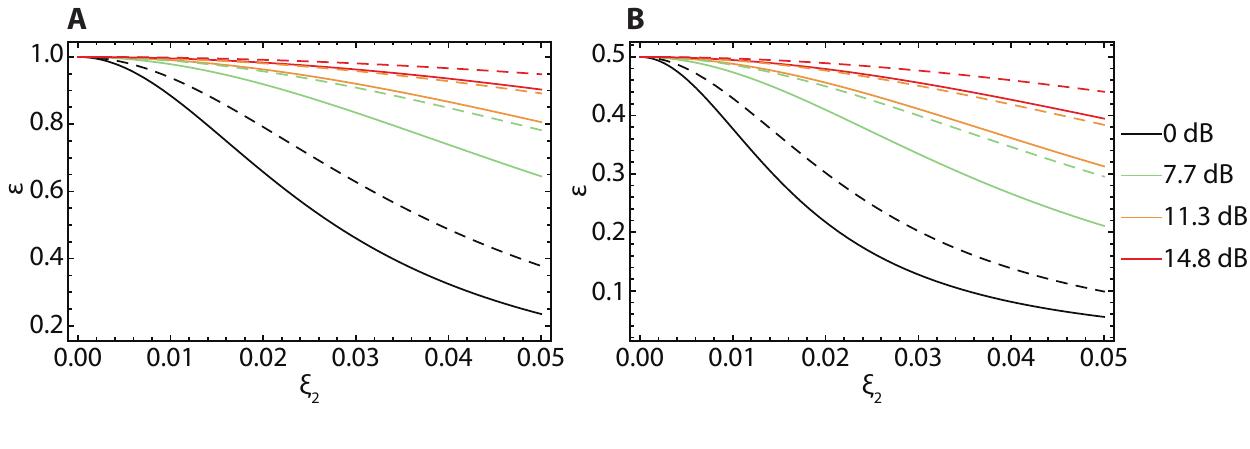}
	\caption{Alice's heralding efficiency. Simulation of Alice's heralding efficiency, as a function of  squeezing parameter $\xi_2$ of source $\mathcal{PS}_2$ with various amount of added loss. Solid and dashed lines correspond to values obtained when $\mathcal{PS}_1$ squeezing parameter was set to $\xi_1\approx0.032$ and $\xi_1\approx0.045$, respectively.  ({\bf A}), Simulation corresponds to the case when each channel has no loss and unit detection efficiency. ({\bf B}), Efficiency of A-PA arm is assumed to be $50\%$, while all other arms have $25\%$ efficiency, which approximately corresponds to the experimental parameters.}
	\label{highorders}
\end{figure*}
\end{document}